\documentclass[aps,prl,twocolumn]{revtex4-1}

\usepackage{amsmath}
\usepackage{amsthm}
\usepackage{dcolumn}
\usepackage{bm}
\usepackage{amssymb}
\usepackage{graphicx}
\usepackage[breaklinks=true]{hyperref}
\usepackage{epstopdf}
\usepackage[flushleft]{threeparttable}
\usepackage{textcomp}
\usepackage{color}
\usepackage[table]{xcolor}
\usepackage{tabularx}
\usepackage[normalem]{ulem}
\usepackage{algorithm}
\usepackage{algorithmicx}
\usepackage[noend]{algpseudocode}

\begin{document}

\title{Explaining the emergence of complex networks through log-normal fitness in a Euclidean node similarity space}
\author{Keith Malcolm Smith}
\affiliation{Usher Institute of Population Health Science and Informatics, University of Edinburgh}
\email{k.smith@ed.ac.uk}

\begin{abstract}
Networks of disparate phenomena-- be it the global ecology, human social institutions, within the human brain, or in micro-scale protein interactions--  exhibit broadly consistent architectural features. To explain this, we propose a new theory where link probability is modelled by a log-normal node fitness (surface) factor and a latent Euclidean space-embedded node similarity (depth) factor. Modelling based on this theory considerably outperforms popular power-law fitness and hyperbolic geometry explanations across 110 networks. Importantly, the degree distributions of the model resemble power-laws at small densities and log-normal distributions at larger densities, posing a reconciliatory solution to the long-standing debate on the nature and existence of scale-free networks. Validating this theory, a surface factor inversion approach on an economic world city network and an fMRI connectome results in considerably more geometrically aligned nearest neighbour networks. This establishes new foundations from which to understand, analyse, deconstruct and interpret network phenomena.

\end{abstract}

\maketitle


\section{Introduction}
Theories and models of the emergence of complex networks allow us to gather insights into their potential generative mechanisms \cite{Watts1998, Barabasi1999a}. The seminal prototype of network models is the Erd\"os-R\'enyi (ER) random graph where all links have equal probability, $p$, of appearing in the graph. A realisation of this random graph is generated by assigning uniformly random values to all node pairs and substantiating the existence of those links whose values lie above the probability threshold, $p$ \cite{Erdos1959}. For a large enough number of nodes, each distinct graph topology (i.e. graph isomorphism class) has roughly equal probability of appearing from this model \cite{Bollobas1998}. Yet, the topological characteristics of real-world networks substantially and consistently deviate from ER random graphs \cite{Newman2006a}, telling us that real-world networks occupy a relatively small and highly uncommon set of graph isomorphism classes. Subsequently, a proliferation of network models have been developed in attempts to understand or reproduce common real-world topological characteristics. 

We can broadly classify network models either as being generative or non-generative. Non-generative models such as configuration models \cite{Sneppen2002, Newman2006a}, stochastic block models \cite{Holland1983},  and complex hierarchy models \cite{Smith2017a} attempt to target or emulate real-world network properties, focused on practical issues such providing null models of specific network properties. Generative models, on the other hand, seek to derive complex network-like topologies from proposed generative mechanisms, the aim of which is to provide plausible physical explanations for the non-arbitrary topological features found in real-world networks from first principles. A popular branch of generative modelling derives from the theory of preferential attachment, where new nodes entering the network have greater probability of linking to nodes with greater numbers of existing links. Such mechanisms have been shown to generate scale-free degree distributions, which have been observed also in some real-world networks \cite{Barabasi1999a}. 
It has also been shown that scale-free networks can instead develop from power-law node `intrinsic fitness', where each node has a probability of forming connections according to a power-law distribution \cite{Caldarelli2002}. 

There is public disagreement among network scientists about how common scale-free degree distributions really are in networks \cite{Holme2019}. Recent work analysing what kinds of distributions best fit degree distributions from a corpus of hundreds of real-world networks suggested that power-law degree distributions accounted for less than 5\% of the corpus, while fitting log-normal distributions achieved equivalent or better results for 88\% \cite{Broido2019}. This quickly generated counter-arguments from scale-free network proponents \cite{Holme2019}. Foremost of which was a work stating that a broader classification of what constituted a scale-free network was required, namely that power-laws need only be present in the right-tail of the degree distribution, rather than the whole distribution (denoted as pure power-laws), for the network to be classified as scale-free \cite{Voitalov2019}. Indeed, it has been known for some time that pure power-law degree distributions are necessarily only found in sparse networks \cite{DelGenio2011}.

One part of the current work demonstrates that the log-normal distribution may be the key to reconciling these viewpoints. First of all, we argue that distributions of abilities or tendencies, such as those proposed in the idea of intrinsic fitness, tend to be log-normal rather than power-law \cite{Limpert2017}. Secondly, the right tails of log-normal distributions approximate power-laws \cite{Mitzenmacher2004}, satisfying the previously mentioned more relaxed definition of scale-free \cite{Voitalov2019}. Thirdly, using modelling we seek to establish if log-normal fitness creates power-law degree distributions at sparse densities and log-normal degree distributions in more dense networks. 

Another branch of generative models considers nodes existing in a latent space and connections occurring where those nodes are close together in the space. The idea that nodes which are similar to each other are more likely to form connections, otherwise described as homophily, is intuitively sensible. By extension, this has led to the theory that some latent space of node similarities underlies the development of network structure \cite{Hoff2002}. A prototype of this approach can be seen as the random geometric graph, where nodes are random samples of an $n$-dimensional Euclidean space and where links form between the closest samples \cite{Dall2002}. This model has some relevant properties to real world networks such as a high modularity and clustering, but does not display the degree heterogeneity implicated by hub nodes typical of complex networks. Further to this, Serrano \textit{et al.} proposed an elegant hyperbolic geometric model where nodes randomly sampled on the unit circle were attached geometrically with constraints for the expected degree distribution of the network \cite{Serrano2008, Allard2017}. Utilising this model, it was then proposed that a trade-off of popularity and similarity was an alternative explanation of network evolution \cite{Papadopoulos2012}. Although this combination of `popularity' and `similarity' is an attractive proposition, and one that will be echoed in the theory of this paper, these works do not provide an explanation for how the degree distributions of complex networks themselves arise.

The literature suggests two major themes in explaining the emergence of complex networks: i) heavy-tailed node fitness-- an individual aspect describing general potentials of nodes for interactivity and ii) homophily-- a pair-wise aspect describing the suitability of pairs of nodes for making links. These here are combined in a new theory, called surface-depth theory, which proposes to model link probability using factors of log-normal fitness (the surface factor) and node similarity embedded in a high dimensional Euclidean space (the depth factor). We rigorously test our theory against prevailing theories of power-law distributions and hyperbolic geometry across over 100 real world networks, showing that our theory significantly and consistently achieves much greater accuracy in emulating real world network topologies. We then describe an application of this theory for recovering the depth factor of weighted complex networks and validate this on pertinent economic and brain networks.

\section{Theory}
\begin{figure*}[!tb]
	\centering
	\includegraphics[trim = 0 20 0 0,clip,scale=.3]{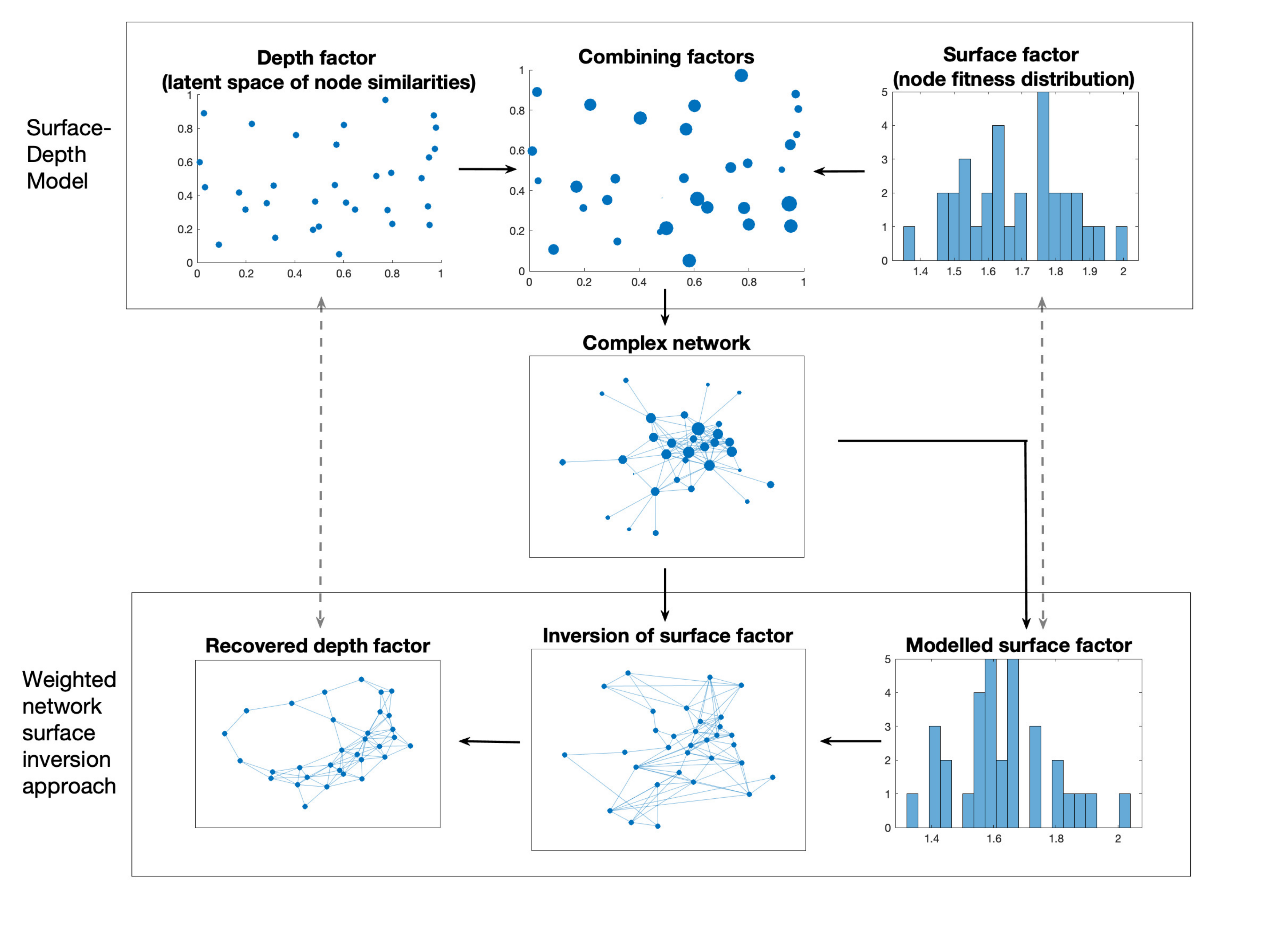}
	\caption{Illustration of surface-depth model and surface inversion approach. The surface-depth model is composed of a depth factor where node similarities are determined by a number of unknown independent latent variables equivalent to dimensions in a Euclidean space (top left) and a surface factor of log-normal node fitness (top right). Each node has a position in the Euclidean space and a node fitness value (top centre). These are combined to constitute the probability of links existing in a complex network (centre). Assuming this to be true, for a weighted network we can model the surface factor (bottom right), divide the network weights appropriately with respect to the surface factor (bottom centre) to obtain an approximation of the depth factor (bottom left).}
	\label{illustration}
\end{figure*}

In the following we combine a number of key existing ideas in the network science literature with novel insights to produce a coherent and simple theory of how complex networks develop their characteristic topologies. 
To aid the reader, an illustration of the different parts of the theory is provided in Fig \ref{illustration}.

\subsection{Surface factor}
Let $\mathcal{V} = \{1,\dots,n\}$ be a set of nodes representative of individual components of a network. Then, suppose that these components have individual tendencies to make links to the other components. Consider in social networks that the tendencies of people to make new friends is the result of a number of psychological variables-- such as extroversion and charisma-- which are general attributes held by individuals. In economics, more open and wealthy countries are more likely to make stronger international ties and have the capacity to maintain more ties. For an example in biology, recent computational experiments indicate plausibility that gene-expression (which influences the concentration of proteins within cells) may aid in the formation of protein-protein interaction networks \cite{Klein2020}. In each case, the collection of tendencies to make links of each node will form some kind of distribution. Whether and what generality of distribution type is possible across such disparate phenomena is a necessary consideration for a universal approach to generative modelling of networks.

Work on understanding the emergence of power-laws in the tails of degree distributions has gravitated towards power-laws themselves as the distribution of such tendencies, referred to as `scale-free node fitness' \cite{Caldarelli2002}. Power-laws tend to crop up in relationships between variables such as in allometry or in dimensions of cities \cite{West2017}, although caution is widely advised in postulating such relationships from observation \cite{Stumpf2012}. In most cases, however, empirical evidence suggests singular variables consist of a large bell shaped concentration of values with a heavy right tail and are well suited to modelling with the log-normal distribution \cite{Limpert2017}. This, in turn, suggests that such variables come from the product of more than one independent random variable, since the product of independent positive random variables tends to the log-normal distribution (via the central limit theorem in the log-scale). Note, a log-normal distribution is typically defined as the distribution resulting from a normally distributed variable as the argument of the exponential function, $s = exp(x)$ where $x\sim N(\mu,\sigma)$. Then, we propose to model the tendency of components to make links as a variable distributed log-normally, $s \sim LogN(\mu,\sigma)$. This is particularly promising given that recent evidence suggests most observed degree distributions of complex networks appear better approximated by log-normal distributions than power-laws \cite{Broido2019}.

Moreover, it is known that the tail of the log-normal distribution resembles a power-law \cite{Mitzenmacher2004}, i.e. a straight line on a log-log plot. The log of a log-normally distributed variable, $x$, is normally distributed, $y = ln(x)\sim N(\mu,\sigma)$, while the log of the probability density function of this normal distribution is a quadratic in $y-\mu$,

\begin{align}
\ln{\left(\text{pdf}(y)\right)} &= \ln{\left(\frac{1}{\sigma\sqrt{2\pi}}e^{-\frac{1}{2}(\frac{y-\mu}{\sigma})^2}\right)}\\
	           &= -\ln{\left(\sigma\sqrt{2\pi}\right)} -\frac{1}{2}\left(\frac{y-\mu}{\sigma}\right)^2.\\
\end{align}
Then the rate of change of this is linear in $y-\mu$ and as the distribution moves further from the mean the fractional change in increase from one point to the next (i.e. $(y_{i+1}$-$y_i)/y_{i+1}$) decreases and the plot tends to a straight line.

Now, we relate to the variable $s$ as the surface factor of the network, since it does not really help to describe why any two nodes are connected together beyond that either or both have strong or weak tendencies to make connections. We could consider whether such tendencies are additive or multiplicative for pairs of nodes, i.e. is the combined tendency of $s_{i}$ and $s_{j}$ $(s_{i} + s_{j})$ or $s_{i}s_{j}$? This is not of immediate importance since the product of two log-normally distributed variables is log-normal, while the addition of two log-normally distributed variables, $x$ and $y$, with the same parameters $\mu$ and $\sigma$ is approximated by the log-normal distribution $x+y\approx z\sim LN(\hat{\mu},\hat{\sigma})$, where
\begin{equation}
\hat{\sigma}^2  =\ln((e^{\sigma^2}+1)/2)
\end{equation}
and 
\begin{equation}
\hat{\mu}=\mu+\ln(2) + (\sigma^2 - \hat{\sigma}^2)/2,
\end{equation}
as described in \cite{Marlow1967}. However, we are concerned primarily with the effect this factor has on the degrees of the network rather than on individual links. In this case, the sum turns out to be more tractable. Consider,
\begin{align}
u_i 	&= \sum_{j\neq i}(s_{i} + s_{j}) \\
	&= (n-1)s_{i} + \left(\sum_{i=1}^{n}s_{j}\right)-s_{i}\\
	&= (n-2)s_{i} + \sum_{i=1}^{n}s_{j}\\
	&= As_{i} + B,
\end{align}
where $A = n-2$ and $B = \sum_{i=1}^{n}s_j$. This is precisely linear in $s_i$, noting that $A$ and $B$ are exactly the same for all $i$. On the other hand,
\begin{align}
v_{i} 	&= \sum_{j\neq i}s_{i}s_{j}\\
	&= s_{i}\sum_{j\neq i}s_{j},
\end{align}
and so there is no such exact linear relationship with $s_{i}$. We could only say that it is approximately $Bs_{i}$ for large enough $n$ and small enough $s_{i}$. Since the sum is more practical for our purposes, we shall here stick with $s_{i} + s_{j}$ as the surface factor for the existence probability of link $(i,j)$.


Note, for the log-normal distribution, we can arbitrarily fix $\mu$ and allow the shape parameter $\sigma$ to vary to produce the different shapes of the distribution, thus essentially, the surface factor has a single parameter, $\sigma$.

\subsection{Depth factor}
Below this surface, we follow the homophily principle by assuming that there are similarities between components which make it more likely for connections to occur between them. In this way, we incorporate the idea of latent spaces encoding similarities between nodes \cite{Hoff2002}. Thus, we suppose that components are distinguishable by some number, $q$, of independent latent variables, $x_{1},x_{2},\dots,x_{q}$. Then, the similarity of nodes $i$ and $j$ across these variables can be described by some inverse distance function (to be consistent with the surface factor `closer' nodes should attain larger values)
\begin{equation}
d_{ij} = f(x_{1}(i),x_{1}(j),x_2(i),x_{2}(j),\dots,x_q(i),x_{q}(j)).
\end{equation}
A very obvious and important consideration of such latent variables is simply the geometry within which the components are set. If two components are proximal to one another, it stands to reason they are more likely to share a link than to share links with components which are further away, disregarding other variables. It is important to point out that latent variables could also be categorical. For instance, in a social network, people who belong to the same club, $A$ say, are more likely to be linked than to others in another club, $B$. 

The geometry of the latent space is an important consideration. Serrano et al. \cite{Serrano2008} developed a latent space model in hyperbolic geometry. Nodes were place on the unit disc (equivalent to the latent space of the model), parameterised by the angle to some arbitrary axis, while the degree distribution of the network was used to parameterise the radius of the node on the disc. While an elegant model, choosing the unit circle as the latent space is problematic as it restricts the dimensionality of the space. 

For our modelling, we need a description of the properties of the latent variables, $x_{i}$. We know that geometry is a key consideration of networks, and thus we have up to three variables which can be approximated using a random geometric graph where coordinates are chosen uniformly at random over the interval $[0,1]$. For simplicity we shall prescribe all variables as independent and identically distributed (i.i.d.), thus we shall simply model similarities between nodes as distances of a random geometric graph in $q$ dimensions. Of course, it is likely that different variables will have different distributive properties in reality, but, as we shall demonstrate, this simple assumption actually works quite well in practice for modelling a diverse range of complex networks. Taking into account that smaller distances should indicate greater probability of attachment, we have, for each link, a depth factor of
\begin{equation}\label{distance}
d_{ij} =  \exp\left(-\sqrt{\sum_{k = 1}^{q}(x_{ik}-x_{jk})^{2}}\right)
\end{equation}
for each $x_{i} \sim U(0,1)$ and independent.

One important detail of i.i.d. latent variables is that the limit of the distribution of their sum as $q\rightarrow\infty$ is a normal distribution, by the central limit theorem. This extends to Euclidean distances between samples: take two randomly sampled points in $q$-dimensional space, $\mathbf{x} = \{x_{1},x_{2},\dots,x_{q}\}$ and $\mathbf{y} = \{y_{1},y_{2},\dots,y_{q}\}$ with each $x_{i},y_{i}\sim U(0,1)$. Then let

\begin{equation}
z_{i} = (x_{i}-y_{i})^2,
\end{equation}
so that each $z_i$ is also i.i.d and, by the central limit theorem, $\sum_{i=1}^{q}z_i$ has a normal distribution in the limit as $q\rightarrow\infty$. From the delta method \cite{Doob1935}, this holds also for functions of the distribution such as the square root-- $\sqrt{\sum_{i=1}^{q}z_i}$-- which is just the Euclidean distance between $\mathbf{x}$ and $\mathbf{y}$ and this further extends to equation \eqref{distance}. This property will be of use later in attempts to invert the surface factor of observed networks.

\subsection{Combining factors}
From the above, the probability of a connection being established between nodes $i$ and $j$ of a network is proportional to both the similarity of the nodes (depth factor) and the combined fitness of the nodes (surface factor), giving
\begin{equation}
p_{ij} \sim d_{ij}(s_{i} + s_{j}).
\end{equation}

Assuming that these are the only considerations of the probability of existence of a link, we can take the weights of links in our network as
\begin{equation}\label{modelweight}
w_{ij} = d_{ij}(s_{i} + s_{j})
\end{equation}
up to linearity. For a complex binary network with $m$ links, we can then, for example, take the $m$ largest weights as extant, use a nearest neighbours connectivity approach \cite{Eppstein1997}, or use a combination of the two to specify the exact number of links while ensuring there are no isolated nodes. The only parameters of this model are the number of dimensions of the depth factor, $q$, and the shape parameter for the log-normal distribution of the surface factor, $\sigma$ and, for a network, $G$, with $n$ nodes and $m$ links, we can describe its surface-depth model as $G_{\text{s-d}}(q,\sigma)$. Note, we intentionally avoid normalising weights to provide exact formula for $p_{ij}$, because we wish to model networks using the same number of nodes and links to avoid the confounding effects of network size and density on network metrics.



\subsubsection{Estimating the surface factor in a weighted network}
Given the above theory, it would be of high interest to uncover the depth factor of real networks as this would help to determine and analyse the similarity structure of nodes beyond the somewhat confounding tendencies for attachment. However, recovering the depth factor of sparse binary networks poses a very challenging problem, as it would seem intractable to determine which links are stronger to a given node than any other from the binary links. What we can do, however is to apply our methods to weighted networks by assuming that the weights of the network are approximately linearly proportional to the underlying link probabilities of the network. This is motivated by the fact that, for example, thresholded functional brain networks display the consistent topological characteristics of binary real world networks \cite{Bullmore2009}.

We saw that distances in Euclidean space have a distribution tending to normal as $q\rightarrow\infty$, and thus approximate the normal distribution for large $q$. Importantly, the normal distribution is a symmetric distribution with 0 skewness. On the other hand, degree distributions of real world networks and those coming from our model are right-skewed (at least for densities $d<0.5$, relevant to most real-world networks). We must presume then, that if our model holds, the majority of this skewness is attributed to the surface factor of the network, while the distribution of depth factor weights has minimal skew. Therefore, we propose here an optimisation algorithm to determine an estimate of the log-normal surface factor of a network by minimising the skewness of network weights after inverting estimated surface factors determined by an array of log-normal distributions. In this case, the argument of the minimisation is the shape parameter $\sigma$ of the log-normal distribution. Supplementary material section i.c demonstrates i) that distances between random samples in an $q$-Euclidean geometric space have highly symmetric distributions even for fairly small $q$, and ii) simulation experiments showing correlations between the real and estimated depth factor weights are inversely related to skewness. Note, without knowledge of the degree distribution of the hypothetical depth factor, we are left with the practical assumption that the ranks of the $n$ random samples of the log-normal distribution align with the ranks of the weighted degrees of the given weighted network.

\section{Materials \& Methods}
Here, we detail the data used in our studies; the details of our modelling approach for real-world networks, alongside the tests and comparisons conducted; and the details of the surface factor optimisation algorithm. For methodological details of more basic exploratory experiments on the model, see Section I of the supplementary material.

\subsection{Real-world network data}
Two datasets of networks were used for the modelling experiments. The first consisted of 25 networks taken from the network repository across different domains \cite{nr2015}. This consisted of eight social networks-- karate club, hi-tech firm, dolphins, wikivote, Hamsterster, Enron email, Dublin contact, and Uni email; six biological networks-- mouse brain, macaque cortex, c elegans metabolism, mouse, plant, and yeast proteins; three ecological networks-- Everglades, Mangwet and Florida; three infrastructure networks-- US airports, euroroads and power grid; and three economic networks-- global city network (binarised at 20\% density), US transactions 1979 commodities  and industries. Many of these were classic benchmark networks. 

The second network dataset was the corpus used in \cite{Ghasemian2018}. Of this dataset, we looked at the 184 static networks and, for the sake of computational time, chose to look only at those between 20 and 500 nodes in size. Further, we discarded bipartite networks as these have 0 clustering and thus obviously need a different depth factor consideration than the random geometric graph which has a large clustering coefficient. This provided a final count of 85 networks.

For the surface inversion examples, we used two well-established weighted networks. The first is the world city network, available from the Globalisation and World Cities research network \cite{Taylor2001,Taylor1999}, constructed using relationships of producer service firms at the forefront of economic influence within each city. Here, each link weight is the sum over service firms of the product of the size the service firm's offices in the two locations, normalised by the value of the maximum possible linkage in the network. In this way it relates how similar the economies of the cities are while having bias towards strength of the economy in the city. Full details are available in \cite{Taylor2001}.

The second was the fairly sparse (link density of 0.0917) weighted group average fMRI network available freely from the brain connectivity toolbox \cite{Rubinov2010}, the foremost resource for brain network analysis algorithms. This fMRI network was derived from a group of 27 healthy individuals. Grey matter was parcellated into 638 regions and the Blood Oxygen-Level Dependent (BOLD) time series was derived for each region. From these, Pearson's correlations of the time-series between pairs of regions were computed and normalised using the Fisher transform. The average values across the 27 individuals were then taken. For full details, see \cite{Crossley2013}.

\subsection{Modelling real-world networks}
For a given network, we found optimal parameters of the surface depth-model based on the Root Mean Squared Error (RMSE) of topological network metrics. We compared our model against two popular existing theories of power-law fitness and hyperbolic geometry. These could be easily incorporated into our analysis by a switching of factors (switching log-normal for power-law in the surface factor and switching Euclidean geometry for spherical geometry in the depth factor). The details are described below.

Five topological network metrics were chosen on which to base the optimisation of the model to a real world network. These were the clustering coefficient, $C$, global efficiency \cite{Latora2001}, $E$, normalised degree variance \cite{Smith2018b}, $V$, modularity based on the Louvain algorithm \cite{Blondel2008}, $Q$, and assortativity \cite{Newman2002}, $r$. Each metric was chosen on the basis that i) it covered a distinctly formulated topological aspect, and ii) its value was appropriately normalised with maximum possible magnitude of 1 so that the minimisation was not evidently biased to any particular index. This kind of minimisation has been previously used in e.g. \cite{Betzel2016,Topirceanu2018}. We assumed that for a node to exist in a sparse binary network, it would be required to be connected within it-- consider that isolated nodes could exist in a system without the knowledge of the network constructor. Thus models (with the same number of nodes as their corresponding real-world networks) were ensured to have all nodes with at least degree 1 by including the nearest neighbours for each node. The rest of the links were then selected simply from the links with highest weights across all model weights until the number of links matched the real network.

After network metrics were computed for each generated model, the RMSE over all metrics between the real-world network and its model
\begin{equation}
RMSE = \sqrt{\frac{1}{T}\sum_{i=1}^{T}(M_{i}-\hat{M}_{i})^2}
\end{equation}
was computed, where, each $M_{i}$ is the value of one of the five metrics defined above (arbitrarily) for the real-world network and $\hat{M}_{i}$ is the corresponding value of that metric for the surface-depth model. In our case, then $T = 5$-- being the five metrics $C,E,V,Q,$ and $r$. The RMSE was used for optimising the model by searching for the model parameters which produced the minimum RMSE. This optimisation was implemented using the following algorithm:
\begin{algorithm}[H]
\caption{Modelling a network}\label{modelling}
	\begin{algorithmic}[1]
	\State Compute metrics $C$, $E$, $V$, $Q$ and $r$ of network $G$
	\For{$q\in\{1,2,\dots,10\}$}
	\State Compute 20 realisations, $G_{s\text{-}d}(q,\sigma)$, of model with the same size and density as $G$ with $\sigma$ ranging from 0.05 up to 1 in steps of 0.05
	\State Compute $C$, $E$, $V$ $Q$ and $r$ of each of these models and take the mean over realisations for each
	\State Compute the RMSE between metrics of $G$ and metric means of $G_{s\text{-}d}(q,\sigma)$
	\State Take $\sigma'$ as the $\sigma$ parameter of minimum RMSE model
	\State Compute 20 realisations of each surface-depth model with $\sigma$ within 0.05 of $\sigma'$ in steps of 0.01
	\State Take the model with the minimum RMSE value from this step as the minimum for the model with $q$ dimensions	
	\EndFor
	\State The minimum across $q$ of the minimum RMSEs across $\sigma$ is then taken as the model of best fit to G
	\end{algorithmic}
\end{algorithm}

Importantly, it is not expected that the discretisation of the surface factor parameter causes any problems here. It is reasonable to assume in this instance that there are no local minima that would confound the optimization because of the discretisation, since the distributions of the surface-factors are smooth, the right-skew of the distributions are monotonic functions (increasing with log-normal and decreasing with power-law) of the parameters, and the distributions themselves have only global maxima and minima. Note also, we took a maximum of $q =10$ arbitrarily to save on time as we assume the topological properties of the model are asymptotic with $q$, as demonstrated in the supplementary material Section I.A. Figure C in Section II of the supplementary material plots the index values of 10 networks and their models alongside results obtained for models utilising surface and depth factors separately, illustrating how the model adapts to each network.

We compared this model against competing theories of power-law fitness \cite{Caldarelli2002} and hyperbolic geometry (alongside higher dimensional spherical surface geometries) \cite{Serrano2008}. The same algorithm was used for power-law fitness and spherical surface geometry by substituting the log-normal parameter, $\sigma\in[0,1]$, for a power-law parameter, $\gamma\in[2,3]$ (the interval within which most scale-free networks are found to follow), and by substituting $q$-dimensional Euclidean geometry for $q$-dimensional spherical surface geometry, respectively. 

For power-law fitness, the link weights were computed as:
\begin{equation}
d_{ij}(s_{i} + s_{j})
\end{equation}
with $s_{i}$ sampled randomly from a power-law distribution with parameter $\gamma$. Again, $\gamma$ was first checked in steps of $0.05$ in the interval $[2,3]$ in the first stage of the Algorithm \ref{modelling} and then steps of $0.01$ in the second stage.
 
For spherical surface geometry, random samples of a $q$-dimensional spherical surface were generated where coordinates for a single sample were obtained from normalising $q$ normally distributed samples and distances between two samples, $x = [x_{1},x_{2},\dots,x_{q}]$ and $y = [y_{1},y_{2},\dots,y_{q}]$,  computed per the formula
\begin{equation}
d(x,y) = acos\left(\sum_{i = 1}^{q}x_{i}y_{i}\right).
\end{equation}
Then the negative of the exponential was taken, following equation \eqref{distance}, and dimensions of spherical geometry were directly substituted for dimensions of Euclidean geometry in Algorithm \ref{modelling}.

Once the best performing parameters for each model were obtained, the RMSE of these best-performing models were compared to assess which model's topology was closest to the real-world network. We also calculated the Spearman correlation coefficient and its $p$-value between each network's best-fit surface factor parameter and depth factor parameter to test the assumption that these parameters should be independent. Next, degree distributions of the log-normal and power-law models were compared against those of the real-world networks by computing the effect sizes (as the normalised $z$-statistic, $z/\sqrt{n^{2}/2n} = z/\sqrt{n/2}$) and $p$-values (the null hypothesis, that the distributions were not different, was rejected in the case that $p\leq0.05$) for the Kolmogorov-Smirnov (KS) two-sample test. This allowed us to assess whether log-normal surface factors could explain the degree distributions of real world networks and how this compared to the popular power-law theory.

\subsection{Surface factor optimisation}
To test the validity of the model in weighted networks, we assessed to what extent an attempted surface inversion of the weights (i.e. dividing the weights in \eqref{modelweight} by $(s_i +s_j)$ to recover $d_{ij}$) outputted weights with stronger geometric qualities and similarity relationships between the nodes.

To do this, we first required a method to best approximate the log-normal distribution which could hypothetically be the distribution of the surface factor. In the Theory section, we noted that random Euclidean distances in a hypercube tend to a normal distribution as the number of dimensions, $q$, tends to infinity. Section III demonstrates that, indeed, even for fairly small $q$, the distribution of distances looks normal and certainly has negligible skewness. Therefore, we proposed to approximate the hypothetical surface factor of a real world weighted network by finding the parameter, $\sigma$, which minimised the skewness after its inversion from the network weights.
Then, for a weighted network with adjacency matrix $\mathbf{W}$ of size $n$ with entries $W_{ij}$, the shape parameter of a log-normal surface factor was estimated, up to two decimal places, by the following algorithm:
\begin{algorithm}[H]
\caption{Estimating the surface factor}\label{surface}
	\begin{algorithmic}[1]
	\For{$\sigma\in\{0.01,0.02,\dots,1\}$}
	\State Compute 1000 realisations, $\{\mathbf{s}_{k}\}_{k=1}^{1000}$, where each $\mathbf{s}_{k}$ is a vector of $n$ samples from log-normal distribution $LN(0.5,\sigma)$
	\State For each $\mathbf{s}_{k}$, order the samples according to the ranks of the weighted degrees of $W$ (e.g. largest sample goes in position $i$ where node $i$ has largest weighted degree) 
	\State For newly arranged $\mathbf{s}_{k}$, compute the surface factor matrix, $\mathbf{S}$, whose entries $S_{ij} = s_{k}(i) + s_{k}(j)$
	\State Compute the depth factor estimation matrix, $\mathbf{D}$, with entries $D_{ij} = W_{ij}/S_{ij}$
	\State Compute the skewness of the non-diagonal entries of $\mathbf{D}$
	\State For each $\sigma$, average the skewness over the 1000 realisations	
	\EndFor
	\State The value of $\sigma$ which achieves minimum average skewness is taken as the optimised estimate of the hypothetical surface factor of $\mathbf{W}$
	\end{algorithmic}
\end{algorithm}
From this, the estimated depth factor matrix $\mathbf{D}$ of the real-world weighted network was obtained as that with the minimum skewness of its entries. To assess the plausibility of $\mathbf{D}$ as a depth factor, we compared the 5-Nearest Neighbour (5NN) graphs of $\mathbf{W}$ and $\mathbf{D}$. Considering that the weighted degrees may be seen as a simpler approximation of any underlying surface factor distribution, without the need to assume log-normality, we also compared our approach with the network of weights obtained by simply dividing weights, $W_{ij}$, by the average of the weighted degrees of the pair of adjacent nodes (i.e. a 'weighted degree inversion'), obtaining the matrix $\mathbf{H}$ with entries
\begin{equation}
	H_{ij} = \frac{2W_{ij}}{\sum_{k=1}^{n}W_{ik} +\sum_{k=1}^{n}W_{jk}}.
\end{equation}
The resulting 5NN graphs of $\mathbf{W}$, $\mathbf{D}$ and $\mathbf{K}$ were assessed in terms of the associations of the nodes. For the world city network, we assessed the proximity of the nearest neighbours on the globe and performed community detection using Louvain's modularity algorithm \cite{Latora2001} to assess to what extent communities were composed of proximal groups of cities. For full details see supplementary Section V. For the fMRI network, we used the provided geometric information of the nodes to assess proximity of nearest neighbours. We also employed community detection and assessed to what extent communities (or modules) were symmetric across the brain (i.e. in what percentage of cases was a right hemisphere region in the same community as a left hemisphere region), as well as to the average distance found within communities. For full details see supplementary Section IV.

\section{Experiments}
Section I.A of the supplementary material provides some initial explorations of the topology of the model covering topological differences between surface-depth models and random geometric graphs and the behaviour of degree distribution with increasing network density. Importantly, we found that surface-depth models have general characteristics associated with real-world networks, such as high clustering coefficient and modularity, high degree heterogeneity, and disassortativity. Furthermore, degree distributions of surface-depth models with $n=1000$ and $q=4$ exhibited power-laws at densities of 1-4\% and log-normal distributions at densities of 4-40\% (specifically, null hypotheses of two-sample KS tests with power-law and log-normal degree distributions could consistently not be rejected at the 5\% level in these cases).

We shall continue with the most pertinent results regarding the modelling of real world networks. We modelled 110 real world binary networks collected from two difference sources. The most accurate surface-depth model was then chosen by optimising for the two model parameters, $\sigma$ and $q$, following Algorithm \ref{modelling}. Note, in each case, the number of nodes and links in the resulting model were kept the same as in the original network. We then did the exact same approach with parameter substitutions for i) power-law fitness instead of log-normal fitness in the surface factor, and, separately, ii) spherical surface geometry for node similarity instead of Euclidean space in the depth factor.

\begin{figure*}[!tb]
	\centering
	\includegraphics[trim = 30 20 0 0,clip,scale=.3]{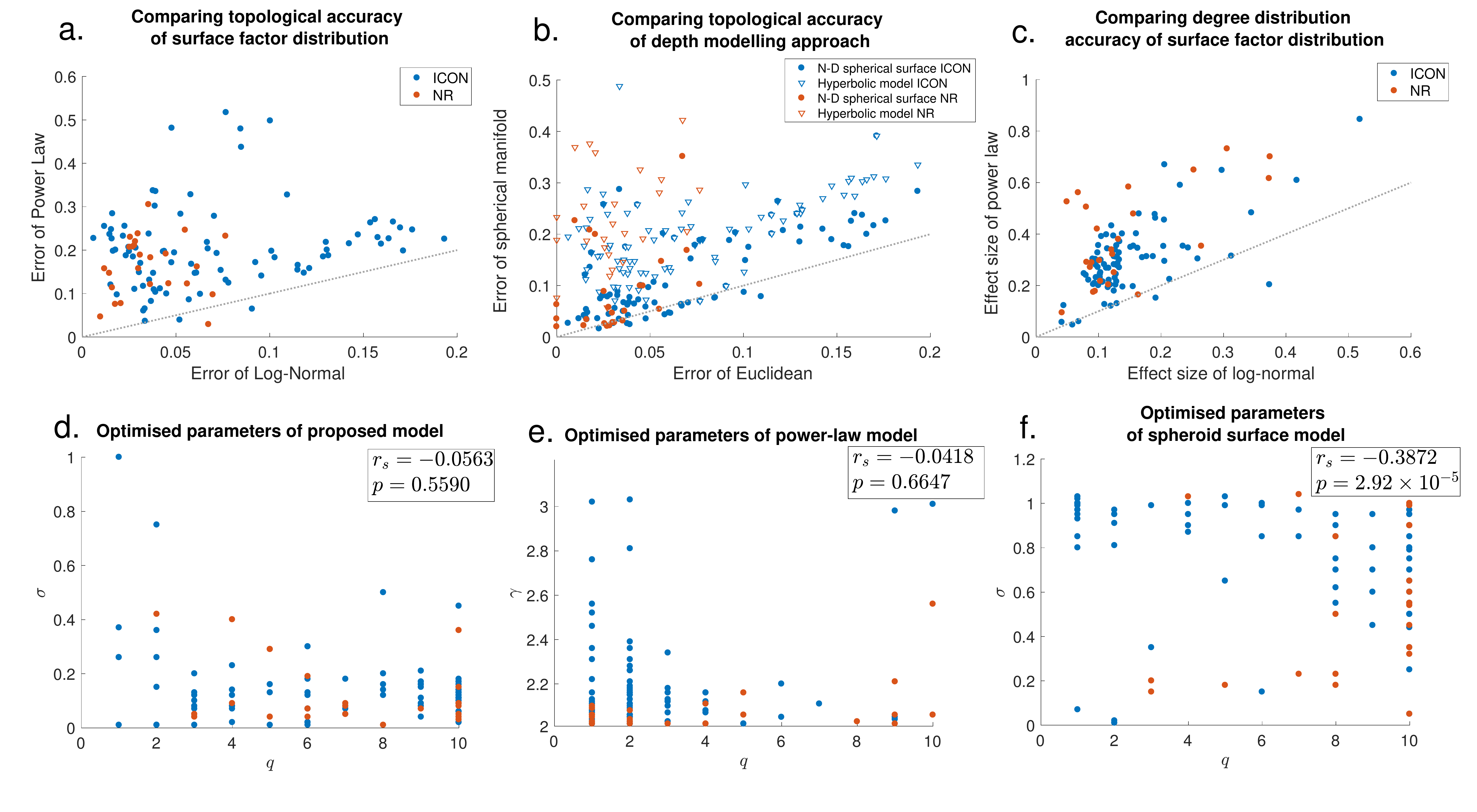}
	\caption{Plots a. and b. show root mean squared errors of the proposed model against power-law attachment spherical surface geometry (including the hyperbolic model), respectively. c. Effect sizes of degree distributions between model and network (log-normal versus power-law attachment). Dotted lines show the line of parity. Plots d., e., and f. show the surface model parameter plotted against the depth model parameter for the proposed theory, power-law attachment theory and spherical surface geometry theory, respectively.}
	\label{figESC}
\end{figure*}

The Root Mean Squared Error (RMSE) in topology of the models for each network-- calculated through five distinct and widely used normalised topological metrics, $C$, $E$, $V$, $Q$ and $r$-- is scatter plotted against RMSE using i) a power-law surface factor and ii) spherical surface depth factor in Fig \ref{figESC} a \& b, respectively. The proposed model clearly outperformed models of theories of both power-law attachment and hyperbolic geometry, with a median RMSE of just 0.0449 compared with 0.1932 and 0.2012 for power-law attachment and hyperbolic geometry, respectively. It also clearly outperformed general $q$-dimensional spherical surface geometry with a median RMSE of 0.0813. In fact, RMSE was smaller in the proposed model than hyperbolic geometry in 99.09\% of networks, power-law fitness in 97.27\% of networks and general spherical surface geometry in 80\% of networks studied.  Furthermore, the average sizes of RMSE were a remarkable 293.4\%, 287.5\% and 170.4\% the size of the proposed model for hyperbolic geometry, power-law fitness and general spherical surface geometry models, respectively.

We then tested to see whether any correlation or anti-correlation was established between the optimised parameters, $q$ and $\sigma$, of the model. The existence of any significant correlation would indicate that the parameters were not independent and thus would negate the claims of the theory that independent surface and depth factors existed to make up link probability. Scatter plots of $\sigma$ against $q$ for all networks are shown for the proposed model, the power-law attachment model and the general spherical surface model in Fig \ref{figESC} a, b \& c, respectively. Spearman's correlation coefficient, $r_{s}$, was used to assess levels of correlation between $q$ and $\sigma$. There was no correlation found between $\sigma$ and $q$ of the proposed theory's model ($r_{s} = -0.0563, p = 0.5590$), validating the independence assumption of surface and depth factors of complex networks. On the other hand, a significant anti-correlation was found between $\sigma$ and $q$ when spherical surface geometry was used ($r_{s} = -0.3872, p = 2.92\times 10^{-5}$), indicating that this model, and the hyperbolic geometry model of which it is a generalisation, was not as appropriate a theoretical foundation for network topology emergence.

\begin{figure*}[!tb]
	\centering
	\includegraphics[trim = 70 70 0 0,clip,scale=.35]{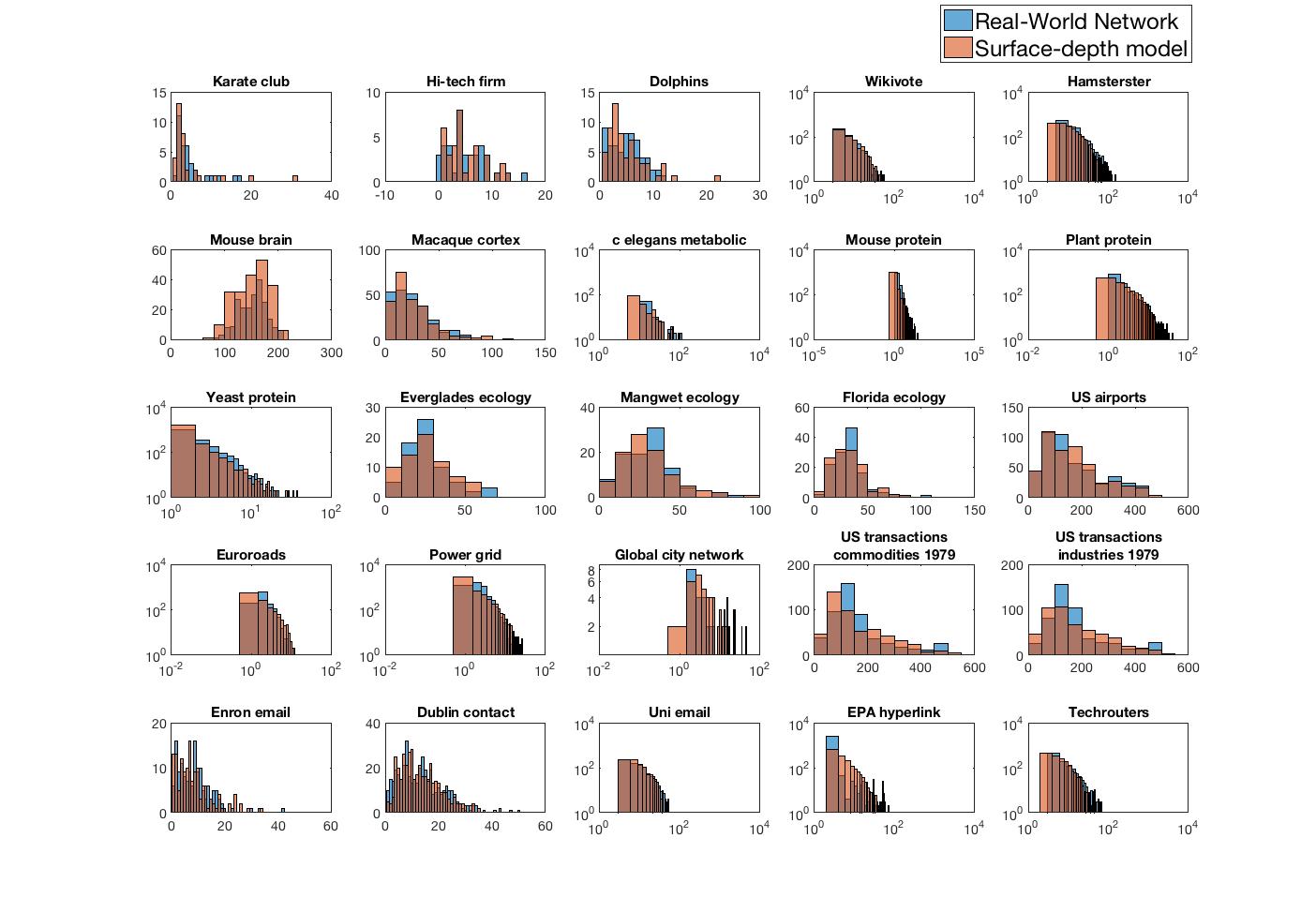}
	\caption{Comparison of the degree distributions between real-world networks and their respective closest fit surface-depth model. These are log-log plots where there is a clear scaling distribution.}
	\label{figModDD}
\end{figure*}

Next, for each real-world network we compared the degree distributions of the best-fit model with the real-world networks using KS two-sample tests. This was done fifty times for each network and median results recorded. Of the 110 networks studied, 68.2\% had no significant median $p$-value, while 81.8\% had no noticeable effect size ($\leq$0.2), with all but one of the remainder (17.27\%) having only small effect sizes ($\in[0.2,0.5]$). Again, these compared very favourably against the power-law fitness model, see Fig \ref{figESC} c. 
Indeed, the average effect size of the power-law model was 225.7\% that of the average log-normal model. Fig \ref{figModDD} shows comparisons of the degree distributions of the network repository networks and their best-fit surface-depth models. The similarity between distributions across all networks of various size, density and domain is striking. From these results, the surface-depth model appears as a good candidate for a unifying theory of attachment in complex network topologies, achieving scale-free like distributions in networks at sparse densities and log-normal like distributions in networks of larger densities, as can be seen in real-world networks in \cite{Broido2019} for example.

Interestingly, there was a particular class of networks that proved to have large errors for all models even though their degree distributions were on the whole largely indistinguishable from those of the proposed model. These were food web networks. Looking more closely, it appeared there was an exceptional difference in the clustering coefficients in this case. Median differences for each index across food web networks were as follows: 0.2753 for $C$, 0.0206 for $E$, 0.0593 for $V$, 0.0185 for $r$, and 0.0449 for $Q$. The very low relative clustering in food web networks makes sense since we can expect that it is uncommon for predators of the same prey to hunt one another as well. This suggests that better modelling of the depth factor may help to better capture the information here.


\subsubsection{Depth factor recovery through estimated surface factor inversion}
To probe further whether surface-depth factors could really be observed in real-world networks, we applied depth factor recovery and subsequent analysis of the recovered depth factor's geometric qualities on two important cases of weighted networks: an economic world city network and a group average fMRI functional brain network, as described in the Methods \& Materials. In both cases, we optimised the log-normal distributions of the surface factors following the network weight skewness minimisation Algorithm \ref{surface} in the methods, based on the fact that Euclidean distances in the $q$-dimensional hypercube tends towards the symmetric normal distribution as $q\rightarrow\infty$ by the central limit theorem, and on the observations in supplementary material Section III.
 


\begin{figure*}[!tb]
	\centering
	\includegraphics[trim = 0 0 0 0,clip,scale=.22]{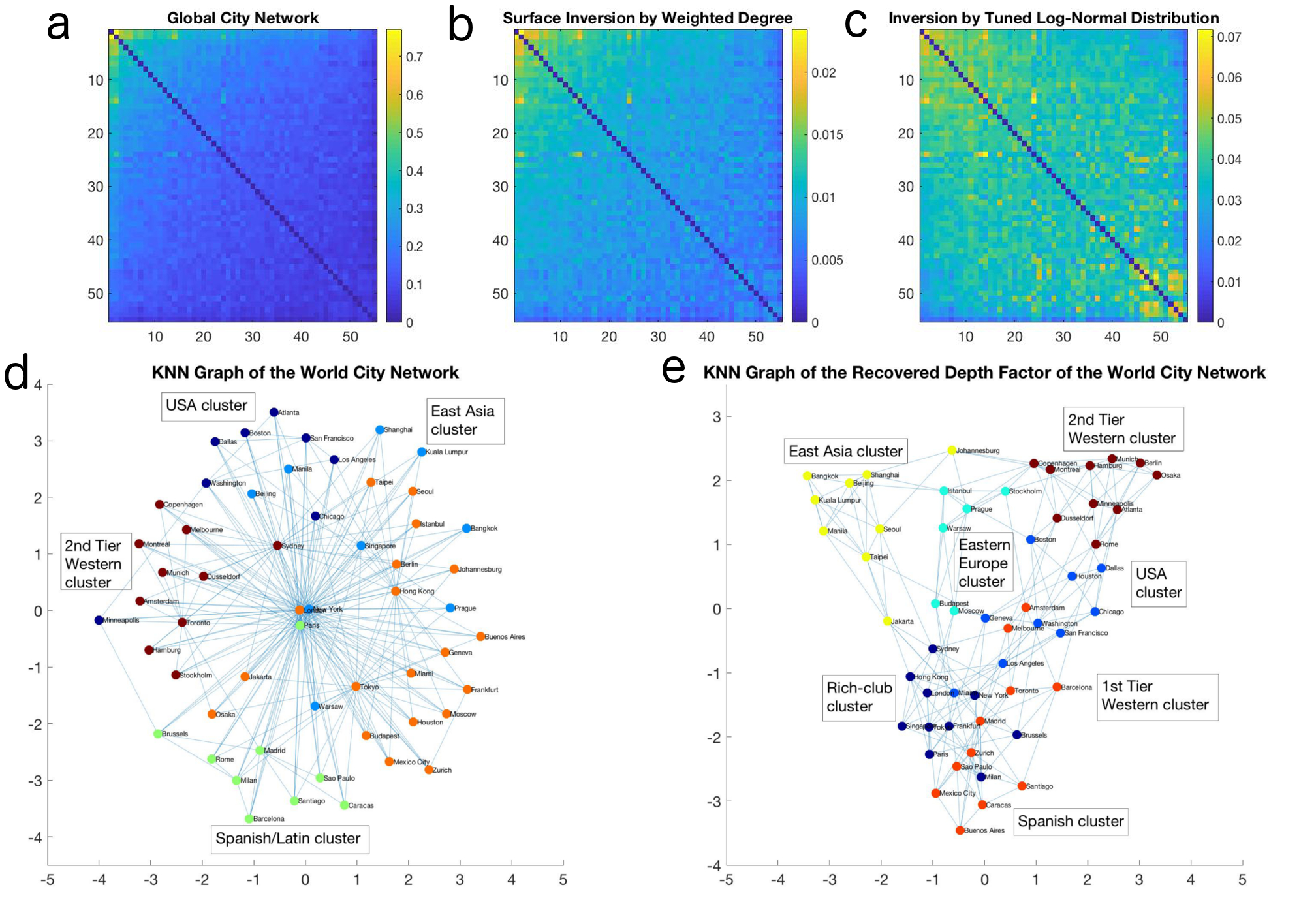}
	\caption{(a) Weighted adjacency matrices (ordered by weighted degree) of the global city network, (b) an estimated depth factor of the network using the weighted degree and (c) an estimated depth factor using a tuned log-normal distribution, respectively. (d) Plot of the five-nearest neighbours graph of the world city network (left) and (e) its recovered depth factor (right) with detected communities shown in different colours. Clusters in the depth factor are observably more distinguishable, whereas relationships between the nodes in the original network are dominated by a few nodes.} 
	\label{gcn_inv}
\end{figure*}

For the global city network, the optimal log-normal distribution was found at $\sigma = 0.59$. K-Nearest Neighbour (KNN) graphs with $K = 5$ were then computed from the global city network and its estimated depth factor. We also compared this with just using the weighted degree distribution as an estimate of the surface factor. Fig \ref{gcn_inv} a, b \& c show the weighted adjacency matrices of the original network and the estimated depth factors from the weighted degree and tuned log-normal distribution surface inversion approaches, respectively.

Modules were computed using Louvain's modularity method \cite{Latora2001}. The 5NN graphs were then plotted using the same force-based algorithm where connected nodes are attracted and non-connected nodes repelled from one another \cite{Fruchterman1991}, Fig \ref{gcn_inv} d \& e. Remarkably, surface inversion of the hub-centric world city network produced a highly modular network with geometric qualities. On inspection, spaces within the network layout were notable by their global proximity and cultural ties. We analysed this statistically in the case of global proximity. Section V of the supplementary material contains these details alongside tables of the five nearest neighbours of each city for each approach. Of these, 180 (65.45\%) were found to be proximal on the globe (either being in the same continent or otherwise geographically close) for the tuned log-normal inversion compared to 50.55\% for the degree-based inversion and just 37.82\% for the original network. Furthermore, the five cities with greatest weighted degree (London, New York, Paris, Tokyo and Hong Kong) appeared in just 10.56\% of the tuned log-normal inversion compared with 76.64\% of the nearest neighbours in the original network and 46.18\% in the degree-based inversion, with 9.27\% being that expected by random chance. In addition, 52 of the 55 cities were found within the 5 nearest neighbours of all cities in the tuned log-normal inversion approach, whereas this number was just 15 for the original network and 38 for the degree-based inversion. All in all, the tuned log-normal inversion provided a remarkably more geometrically congruent network, with a clear elimination of rich-club-style \cite{Colizza2006}  bias in nearest neighbours. Some qualitative observations are also worth noting. Barcelona and Madrid were found to be in the same community as all Latin American cities, appealing to their cultural ties, whereas Latin American cities were not even all found in the same community in the original network. Further, Eastern Europe and East Asia both had clearly distinct communities in the recovered depth factor but not so in the original network.

For the fMRI network, the optimal log-normal distribution was found at $\sigma = 0.27$.  The availability of the 3D coordinates of the nodes representing brain regions allowed us to construct a geometric graph for comparison. The sparsity of the network posed a significant confounding factor in this instance as only those links which already existed could be chosen in the resulting 5NN graph. Nonetheless, we considered three measurements of the geometric appropriateness of the resulting depth factor: i) the percentage of overlapping links with the 5NN graph of the geometric network, ii) the proportion of symmetric nodes across brain hemispheres appearing in the same module, and iii) the average largest distance within modules. Details of these analyses are in the supplementary material Section IV. In all cases the estimated depth factor outperformed the original network. The depth factor achieved consistently greater geometric overlap and module symmetry and smaller average largest distance within modules. This again demonstrates the relationship of the estimated depth factor with underlying geometry of the considered networks.

The combined evidence from the world city and fMRI networks provides promising evidence of the real existence of surface and depth factors in complex networks, substantiating the real-world applicability of the proposed theory and opening up new avenues for discovery in weighted network analysis particularly.

\section*{Limitations and future work}
The theory put forward is topologically accurate in modelling most of the complex networks studied here, yet we made no attempt to take into account dynamically changing networks and network evolution. That being said, it would seem that evolution and dynamics of networks could be incorporated in our theory by shifts occurring in surface and depth factors. For instance, a node may take on different values of its latent variables thus changing the nodes to which it is most similar which would result in a change to the links the node makes. Otherwise, the node may increase or decrease its fitness giving it a higher/lower tendency to make connections, again resulting in a dynamic change of the network. New nodes could be assumed to appear somewhere within the latent variable space but with an initially low tendency to make the connections. Such processes could be stochastically encoded.
 
Also there are evident limitations in the modelling of the depth factor. To improve the model's accuracy, new methods would be required for more accurate depth factors and the fusion of different types of latent variables, including categorical variables and variables with different distributions, as well as weighting variables for their importance. The proposal that a depth factor of weight similarities can be extracted has clear implications in terms of geometric deep learning \cite{Bronstein2017}. Along similar lines, a recent study considered using machine learning approaches on a hyperbolic network model \cite{Muscoloni2017}. It seems that such methods can be fairly straightforwardly translated to the geometries of the proposed depth factor and we expect our study will open up interesting future research along these lines. Immediate applications of the theory include surface inversion to other weighted networks and the consideration of this theory to advance efforts in important network problems such as community detection and link prediction.

\subsection*{Acknowledgements}
The author is thankful to Anton Pichler for useful discussions concerning economic networks. This work was supported by Health Data Research UK (MRC ref Mr/S004122/1), which is funded by the UK Medical Research Council, Engineering and Physical Sciences Research Council, Economic and Social Research Council, National Institute for Health Research (England), Chief Scientist Office of the Scottish Government Health and Social Care Directorates, Health and Social Care Research and Development Division (Welsh Government), Public Health Agency (Northern Ireland), British Heart Foundation and Wellcome.

\textbf{Data availability statement}
Datasets used are readily available and as referenced in this article. Code used is freely available at DOI 10.17605/OSF.IO/PMXU7.

\textbf{Author contributions}
KS is the sole author and did all of the presented work.

\textbf{Competing interests}
The author declares no competing interests.

\bibliographystyle{ieeetr}
\bibliography{References}

\end{document}